\documentclass[aps,11pt]{article}
\usepackage{epsfig,sint,macros,cite,mathbbol,graphicx}

\begin{document}
\begin{titlepage}
\begin{flushright}
MKPH-T-10-10
\end{flushright}

\renewcommand{\thefootnote}{\fnsymbol{footnote}}

\vskip 0.5 cm
\begin{center}
  {\Large\bf Renormalization of minimally doubled fermions \\[0.5ex]}
\end{center}
\vskip 1.0cm
\begin{center}
{\large Stefano Capitani$^{a}$\footnote{capitan@kph.uni-mainz.de},
  Michael Creutz$^{a,b}$\footnote{mike@latticeguy.net},
  Johannes
  Weber$^{a}$\footnote{weberj@kph.uni-mainz.de}\footnote{Present
  address: Graduate School of Pure and Applied Physics, Tsukuba
  University, Tsukuba, Ibaraki, Japan}
  and Hartmut Wittig$^{a}$\footnote{wittig@kph.uni-mainz.de}
}
\vskip 1.0cm
$^{a}$\,Institut f\"ur Kernphysik, Becher Weg 45,
University of Mainz, D-55099 Mainz, Germany 
\vskip 0.3cm
$^{b}$\,Physics Department, Brookhaven National Laboratory, Upton,
NY~11973, USA
\vskip 2.5cm
{\bf Abstract}
\vskip 1.0ex
\end{center}

\renewcommand{\thefootnote}{\arabic{footnote}}

\noindent

We investigate the renormalization properties of minimally doubled fermions, 
at one loop in perturbation theory. Our study is based on the 
two particular realizations of Bori\c{c}i-Creutz and Karsten-Wilczek. 
A common feature of both formulations is the breaking of hyper-cubic 
symmetry, which requires that the lattice actions are supplemented 
by suitable counterterms.
We show that three counterterms are required in each case and
determine their coefficients to one loop in perturbation theory. For
both actions we compute the vacuum polarization of the gluon. It is
shown that no power divergences appear and that all contributions
which arise from the breaking of Lorentz symmetry are cancelled by the
counterterms. 
We also derive the conserved vector and axial-vector currents for 
Karsten-Wilczek fermions. Like in the case of the previously studied 
Bori\c{c}i-Creutz action, one obtains simple expressions, 
involving only nearest-neighbour sites. We suggest methods how to fix 
the coefficients of the counterterms non-perturbatively and discuss 
the implications of our findings for practical simulations.

\vfill

\begin{center}
June 2010
\end{center}

\eject

\vfill
\eject

\end{titlepage}

\setcounter{footnote}{0}

\section{Introduction}
\label{sec:intro}
 
Minimally doubled fermions\,\cite{mind:Karsten81,mind:Wilczek87,
mind:Creutz07,mind:Borici07,Cichy:2008gk,Bedaque:2008xs,
Bedaque:2008jm,Buchoff:2008ei,mind:Creutz08,mind:Borici08,
mind:Capitani09,mind:Capitani_lat09,Kimura:2009qe,Kimura:2009di}
preserve an exact chiral symmetry for a degenerate doublet of quarks,
thereby realizing the minimal doubling of fermion species allowed by
the Nielsen-Ninomiya theorem\,\cite{Nini1,Nini2,Nielsen:1981hk}. At
the same time they also remain strictly local, and may thus be
regarded as a cost-effective realization of chiral symmetry at
non-zero lattice spacing, which is particularly suited for simulating
a degenerate light doublet of up and down quarks.

In a previous article \cite{mind:Capitani09} we began an investigation
into the renormalization properties of a particular realization of
minimally doubled fermions, described by the Bori\c{c}i-Creutz action
\cite{mind:Creutz07,mind:Borici07,mind:Creutz08,mind:Borici08}, based
on perturbation theory at one loop. In this
paper\,\footnote{Preliminary results were reported in
\cite{mind:Capitani_lat09}.} we present a similar analysis for another
member of this class of fermions, proposed a long time ago by Wilczek
\cite{mind:Wilczek87}, following previous work of
Karsten\cite{mind:Karsten81}. Here we take the opportunity to revise
and sharpen some of the conclusions of our earlier
article\,\cite{mind:Capitani09}, by interpreting our results in a more
general field-theoretical framework. In particular, what we
interpreted in ref.\,\cite{mind:Capitani09} as a shift in the quark's
four-momentum can, in fact, be absorbed by adding appropriate
counterterms to the lattice actions, so that any momentum shift
disappears in the fully renormalized theory. The main focus of the
present article is on the one-loop structure of minimally doubled
fermions, taking the Bori\c{c}i-Creutz and Karsten-Wilczek actions as
particular examples. We identify the possible counterterms which can
arise and study the consequences that can be inferred from their
presence.

Our findings can be summarized by the following: a consistent
renormalized theory for minimally doubled fermions can be constructed
by fixing the coefficients of three counterterms in the action which
are allowed by the symmetries. In this work we determine these
coefficients in perturbation theory, and discuss possible
renormalization conditions to fix their values non-perturbatively.
One of the principal results, namely the full expressions for the
renormalized actions including all counterterms, can be found in
eqs.\,(\ref{eq:BCfullaction}) and~(\ref{eq:KWfullaction}). We also
present the computation of the vacuum polarization for both particular
formulations. Our calculation demonstrates that radiative corrections
at one loop do not introduce new divergences for this quantity.

This article is organized as follows. After defining the two actions
and the corresponding propagators and vertices in Section
\ref{sec:actions_etc}, we introduce in Section \ref{sec:counterterms}
the necessary counterterms which render the theories consistent under
renormalization. With these tools in hand we fix the coefficients 
of the counterterms which appear in the quark action in
Section\,\ref{sec:fermcts}, compute the matrix elements of quark
bilinears and derive the conserved currents in
Section\,\ref{sec:bilin_cons} and also show that their renormalization
constant is one. In Section \ref{sec:vacpol} we present the
calculation of the vacuum polarization of the gluon, while in Section
\ref{sec:simul} we discuss the implications of our findings for the
practical implementation of minimally doubled fermions in numerical
simulations. Finally, we make some concluding remarks.

\section{Actions, propagators and vertices} 
\label{sec:actions_etc}

In order to make this article self-contained, we recall here the basic
definitions for Bori\c{c}i-Creutz fermions
\cite{mind:Creutz07,mind:Borici07,mind:Creutz08,mind:Borici08} and
also introduce the Karsten-Wilczek
action\,\cite{mind:Karsten81,mind:Wilczek87}. In position space the
lattice action of Bori\c{c}i-Creutz fermions reads
\begin{eqnarray}
& & S^f_{\rm{BC}} = a^4 \sum_{x} \bigg[ \frac{1}{2a} \sum_{\mu=1}^4 \Big[
    \overline{\psi} (x) \, (\gamma_\mu + i\gamma'_\mu) \, 
   U_\mu (x) \, \psi (x + a\widehat{\mu}) \nonumber \\
&& \phantom{S^f = a^4 \sum_{x} \bigg[ \frac{1}{2a} \sum_{\mu}}
-\overline{\psi} (x + a\widehat{\mu}) \, (\gamma_\mu - i\gamma'_\mu) \,
   U_\mu^\dagger (x) \, \psi (x) \Big] +
\overline{\psi}(x) \, \Big(m_0-\frac{2i\Gamma}{a}\Big) \, \psi (x) 
    \bigg] ,
\label{eq:creutz-action-pos}
\end{eqnarray}
where the matrices $\Gamma$ and $\gamma'_\mu$ are defined by
\begin{equation}
  \Gamma = \frac{1}{2} \, \sum_{\mu=1}^4 \gamma_\mu,
  \qquad\gamma'_\mu = \Gamma \gamma_\mu \Gamma = \Gamma - \gamma_\mu ,
\label{eq:Gamma_def}
\end{equation}
with $\Gamma^2=1$. The Karsten-Wilczek action reads\,\footnote{In this
article we will denote 4-vectors with Greek indices and spatial
3-vectors with Latin indices. The temporal component is $\mu=4$.}
\begin{eqnarray}
& & S^f_{\rm{KW}} = a^4 \sum_{x} \bigg[ \frac{1}{2a} \sum_{\mu=1}^4 \Big[
    \overline{\psi} (x) \, (\gamma_\mu -i\gamma_4 \, (1-\delta_{\mu 4}) ) \,  
    U_\mu (x) \, \psi (x + a\widehat{\mu}) 
\label{eq:wilczek-action-pos} \\
&& \phantom{S^f = a^4 \sum_{x} \frac{1}{2a}}
-\overline{\psi} (x + a\widehat{\mu}) \, 
(\gamma_\mu +i\gamma_4 \, (1-\delta_{\mu 4}) ) \, U_\mu^\dagger (x) \, 
  \psi (x) \Big] 
 + \overline{\psi}(x) \, \Big(m_0+\frac{3i\gamma_4}{a}\Big) \, \psi (x) 
    \bigg]. \nonumber 
\end{eqnarray}
We remind the reader that one Dirac spinor, $\psi(x)$, in these
expressions describes a degenerate doublet of quarks. In momentum space 
the free Dirac operator of Bori\c{c}i-Creutz fermions takes the form
\begin{equation}
  {\cal{D}}_{\rm{BC}}(p) = D(p)+\overline{D}(p)-\frac{2i\Gamma}{a}+m_0,
\label{eq:creutz-action}
\end{equation}
with $D(p)$ and $\overline{D}(p)$ given by
\begin{equation}
 D(p)= \frac{i}{a}\sum_{\mu=1}^4 (\gamma_\mu \sin ap_\mu),\qquad
\overline{D}(p)=\frac{i}{a}\sum_{\mu=1}^4 (\gamma'_\mu \cos ap_\mu).
\end{equation}
This action has two doublers, corresponding to the two zeros (Fermi
points) at $ap_1=(0,0,0,0)$ and $ap_2=(\pi/2,\pi/2,\pi/2,\pi/2)$. For
Karsten-Wilczek fermions the free Dirac operator reads
\begin{equation}
  {\cal{D}}_{\rm{KW}}(p) = \frac{i}{a} \sum_{\mu=1}^4 \gamma_\mu
   \sin ap_\mu  
+ \frac{i}{a} \gamma_4 \sum_{k=1}^3 (1-\cos ap_k) + m_0 
= D(p) + \frac{i}{a} \gamma_4 \sum_{k=1}^3 (1-\cos ap_k) + m_0.
\label{eq:wilczek-action}
\end{equation}
In the latter case the term proportional to $\gamma_4$, which is
chirally invariant since it anticommutes with $\gamma_5$, removes 14
of the doublers of the naive fermion action $D(p)$, and only the
doubler whose pole lies entirely in the temporal direction
survives. The Dirac operator ${\cal{D}}_{\rm{KW}}(p)$ of the
Karsten-Wilczek action exhibits then only two Fermi points, which are
located at $ap_1=(0,0,0,0)$ and $ap_2=(0,0,0,\pi)$ and describe --
like for Bori\c{c}i-Creutz fermions -- two degenerate fermion species
of opposite chirality.

In principle, the term proportional to $\gamma_4$ can be multiplied by
some coefficient, $\lambda$, without spoiling the chiral symmetry of
Karsten-Wilczek fermions, or modifying the minimal number of doublers. 
However, it is conceivable that for general values of $\lambda$ the 
transfer matrix does not exist or that the theory presents some other 
kind of problem. In this work we stick to the case $\lambda=1$, 
which is the common choice in the literature.\,\footnote{This coefficient 
$\lambda$ is in many ways similar to the Wilson parameter of Wilson 
fermions, which is commonly set to one in simulations.}

The two actions described here and investigated at length in this
article represent two particular realizations of minimally doubled
fermions, which respect chiral symmetry at any finite lattice spacing,
but are no longer symmetric under the full hyper-cubic group. The
Bori\c{c}i-Creutz action is only symmetric with respect to a subgroup
which preserves the hyper-cubic positive major diagonal. The term
proportional to $\gamma_4$ in the Dirac operator of Karsten-Wilczek
fermions selects, instead, a different particular direction in 
Euclidean space, the temporal axis.

The breaking of hyper-cubic symmetry allows for mixing with operators
of various dimensions. In particular, mixing with lower-dimensional
operators may occur, which implies the appearance of power-divergent
coefficients proportional to $1/a^n$. In the following we will show
that for these actions the breaking of hyper-cubic symmetry indeed
generates linearly divergent counterterms, whose coefficients need to
be determined via some physical condition.

We will see in the remainder of this article that in several instances
the same reasoning applies to both actions considered. However, we
would like to stress here that, despite many similarities, these two
realizations of minimally doubled fermions are not equivalent. 
The ways in which the two actions are constructed, and the 
mechanisms for the removal of doublers, are indeed qualitatively 
different.\,\footnote{There is some freedom in constructing actions of 
the Bori\c{c}i-Creutz or Karsten-Wilczek type. For instance, in the 
latter case one can replace $(1-\cos ap_k)$ with other trigonometric 
functions without spoiling the basic properties of this kind of action, 
especially the fact that there are two Fermi points exactly 
located at $ap_1=(0,0,0,0)$ and $ap_2=(0,0,0,\pi)$.}

Although the distance between the two Fermi points $p_1$ and $p_2$ is
the same in each of the two actions considered (i.e.,
$p_2^2-p_1^2=\pi^2/a^2$), these two actions cannot in fact be
transformed into each other by a 4-dimensional rotation. On the other
hand, in this article we will show that many results turn out to be
qualitatively similar, owing to the common feature of the breaking of 
the hyper-cubic symmetry by a fixed direction in four-dimensional space 
(determined by the two Fermi points).

It is easy to verify that both actions satisfy $\gamma_5$-hermiticity, 
with all the advantages and simplifications that this property implies, 
especially in numerical simulations.
Another interesting observation is that the Dirac operators can be written as
\begin{eqnarray}
D^f_{\rm{BC}} & = & \frac{1}{2} \, \Bigg\{ 
\sum_{\mu=1}^4 \gamma_\mu (\nabla_\mu + \nabla^\ast_\mu) \, 
+ia \sum_{\mu=1}^4 \gamma'_\mu \,\nabla^\ast_\mu \nabla_\mu \Bigg\} , 
\label{eq:bc-covder} \\
D^f_{\rm{KW}} & = & \frac{1}{2} \, \Bigg\{ 
\sum_{\mu=1}^4 \gamma_\mu (\nabla_\mu + \nabla^\ast_\mu) \, 
-ia \gamma_4 \sum_{k=1}^3 \nabla^\ast_k \nabla_k \Bigg\} , 
\label{eq:kw-covder}
\end{eqnarray}
where 
$\nabla_\mu \psi(x) = 
\frac{1}{a}\,[U_\mu(x)\,\psi\,(x+a\widehat{\mu}) - \psi(x)]$ 
is the nearest-neighbour forward covariant derivative, and $\nabla^\ast_\mu$ 
the corresponding backward one. Thus, it becomes apparent that the two 
realizations of minimally doubled fermions bear a close formal 
resemblance to Wilson fermions, i.e.
\begin{equation}
D^f_{\rm{Wilson}}  = \frac{1}{2} \, \Bigg\{ 
\sum_{\mu=1}^4 \gamma_\mu (\nabla_\mu + \nabla^\ast_\mu) \, 
-ar \sum_{\mu=1}^4 \nabla^\ast_\mu \nabla_\mu \Bigg\} .
\end{equation}
Moreover, this demonstrates the presence of dimension-five operators 
in all three cases. For Wilson fermions the dimension-five operator
breaks chiral symmetry, while for minimally doubled fermions chiral symmetry
is preserved at the expense of introducing operators which break
hyper-cubic symmetry.

For the derivation of the propagators and vertices of
Bori\c{c}i-Creutz fermions we refer to \cite{mind:Capitani09}. Here we
only remind that the Bori\c{c}i-Creutz quark propagator can be written
as
\begin{equation}
S(p) = a\,\frac{\displaystyle -i\sum_\mu \Big[ \gamma_\mu \sin ap_\mu 
            -2\,\gamma'_\mu\,\sin^2 ap_\mu/2 \Big] +am_0}{\displaystyle 
   4 \sum_\mu \Big[ \sin^2 ap_\mu/2 + \sin ap_\mu 
      \Big(\sin^2 ap_\mu/2 
      -\half \sum_\nu \sin^2 ap_\nu/2 \Big)\Big]
      +(am_0)^2},
\label{eq:cpropFP0}
\end{equation}
the quark-quark-gluon vertex is given by
\begin{equation}
V_1(p_1,p_2) = - i g_0 \left(\gamma_\mu \cos \frac{a(p_1+p_2)_\mu}{2}
       -\gamma'_\mu \sin \frac{a(p_1+p_2)_\mu}{2} \right) 
\end{equation}
(where $p_1$ and $p_2$ are the incoming and outgoing quark momenta at
the vertex), and the quark-quark-gluon-gluon vertex is
\begin{equation}
V_2(p_1,p_2) = \frac{1}{2} i a g_0^2 \left(\gamma_\mu \sin 
     \frac{a(p_1+p_2)_\mu}{2} +\gamma'_\mu \cos
     \frac{a(p_1+p_2)_\mu}{2} \right) . 
\end{equation}

The fermion propagator of Karsten-Wilczek fermions is obtained by
inverting the Dirac operator of \eq{eq:wilczek-action}, and is given
by
\begin{equation}
S(p) = a\,\frac{\displaystyle -i\sum_{\mu=1}^4 \gamma_\mu \sin ap_\mu 
            -2i\,\gamma_4\,\sum_{k=1}^3 \sin^2 ap_k/2 +am_0}{\displaystyle 
      \sum_{\mu=1}^4 \sin^2 ap_\mu + 4 \sin ap_4 \,\sum_{k=1}^3 \sin^2 ap_k/2
       + 4 \left( \sum_{k=1}^3 \sin^2 ap_k/2 \right) 
             \left( \sum_{l=1}^3 \sin^2 ap_l/2 \right)
      +(am_0)^2}.
\label{eq:propFP0}
\end{equation}
As in the case of Bori\c{c}i-Creutz fermions, but unlike many standard
fermionic discretizations, we find that the denominator of this propagator 
does not possess a simple behavior under all momentum inversions 
(in this case of the fourth direction, so that it amounts to a 
time reversal).

After making the substitution $ap_4 \to \pi + ap_4$, one obtains the
propagator for the fermionic mode associated with the other Fermi
point, the one at $ap=(0,0,0,\pi)$:
\begin{equation}
S(p) = a\,\frac{\displaystyle i\sum_{\mu=1}^4 \gamma'_\mu \sin ap_\mu
            -2i\,\gamma'_4\,\sum_{k=1}^3 \sin^2 ap_k/2 +am_0}{\displaystyle 
      \sum_{\mu=1}^4 \sin^2 ap_\mu - 4 \sin ap_4 \,\sum_{k=1}^3 \sin^2 ap_k/2
       + 4 \left( \sum_{k=1}^3 \sin^2 ap_k/2 \right) 
             \left( \sum_{l=1}^3 \sin^2 ap_l/2 \right)
      +(am_0)^2},
\end{equation}
where we have also introduced a new set of Dirac matrices,
$\gamma'_k=-\gamma_k$ and $\gamma'_4=\gamma_4$. If we now invert the
direction of the four-momentum $p_\mu$ in this expression, the
propagator of \eq{eq:propFP0} is recovered. Since
$\gamma'_5=-\gamma_5$, we conclude that the modes corresponding to the
two Fermi points have, as expected, opposite chirality.
Note that the symmetry $ap_4 \to \pi - ap_4$ exchanges the zeros.

In analogy to Bori\c{c}i-Creutz fermions, we have also derived the
quark-quark-gluon vertex for Karsten-Wilczek fermions, which reads
\begin{equation}
V_1(p_1,p_2) = - i g_0 \left(\gamma_\mu \cos \frac{a(p_1+p_2)_\mu}{2}
       +\gamma_4 \, (1-\delta_{\mu 4}) \, \sin
       \frac{a(p_1+p_2)_\mu}{2} \right) , 
\end{equation}
while the quark-quark-gluon-gluon vertex comes out as
\begin{equation}
V_2(p_1,p_2) = \frac{1}{2} i a g_0^2 \left(\gamma_\mu \sin 
     \frac{a(p_1+p_2)_\mu}{2} -\gamma_4 \, (1-\delta_{\mu 4}) \, \cos
     \frac{a(p_1+p_2)_\mu}{2} \right) .
\label{eq:v2kw}
\end{equation}
The expressions for the vertices can also be easily derived by
comparing the Dirac operator of Karsten and Wilczek with the
Wilson-Dirac operator
\begin{equation}
D_{\rm w}(p) = \frac{1}{a}\sum_\mu \Big\{i\gamma_\mu \sin ap_\mu + r \,
(1-\cos ap_\mu)\Big\} + m_0, 
\label{eq:wilson-action}
\end{equation}
and observing that the hopping terms of these two actions are related
by the replacement $r \leftrightarrow i\gamma_4$, with a further
restriction of the term proportional to $\gamma_4$ to its spatial
components only.

\section{Counterterms}
\label{sec:counterterms}

A key result of this paper is the observation that the actions
discussed in the previous section do not contain all possible allowed
operators that are invariant under the subgroup of the hyper-cubic
group which is left as a symmetry (preserving the positive major
diagonal for the Bori\c{c}i-Creutz case or the temporal axis for
Karsten-Wilczek). We will describe in detail how radiative corrections
generate new contributions whose form is not matched by any of the
terms in the original bare actions. These operators must then be
added as counterterms in order to construct a consistent theory under
renormalization, and this consistency requirement will uniquely
determine the magnitude of their coefficients.

In the following we will consider the massless case, $m_0=0$. Our task
is to construct and add to the bare actions presented in
Section\,\ref{sec:actions_etc} all possible counterterms which are
allowed by the remaining symmetries. For this purpose we consider
operators of dimension four or lower. We first classify them using the
common notation in the continuum and then proceed to specify
convenient lattice representations of these operators.

We begin with the fermionic part of the actions. The presence 
of a conventional chiral symmetry strongly restricts the number 
of possible counterterms. Since they have to anticommute with 
$\gamma_5$, we can restrict the list to those operators 
which contain $\gamma_\mu$. Other Dirac matrices like $1$,
$\gamma_5$, $\gamma_\mu\gamma_5$ and $\sigma_{\mu\nu}$ can be
excluded. The particular way in which symmetry breaking occurs in
Bori\c{c}i-Creutz fermions implies that we are allowed to construct
operators where summations over single indices occur, in addition to
the standard Einstein summation over two indices. Then also operators
containing $\sum_\mu \gamma_\mu = \Gamma$ are permitted.

As a consequence, for Bori\c{c}i-Creutz fermions there is only one
possible counterterm of dimension four, namely
$\overline{\psi}\,\Gamma \sum_\mu D_\mu \psi$ (which amounts to a 
renormalization of the speed of light for the fermions, relative to the 
positive diagonal axis). We can represent it on the lattice by writing it 
in a form similar to that of the hopping terms already present 
in the action. More precisely, we use the gauge invariant expression
\begin{equation}
c_4 (g_0)\,
\frac{1}{2a} \sum_\mu \Big( \overline{\psi} (x) 
\, \Gamma \, U_\mu (x) \, \psi (x + a\widehat{\mu}) 
-\overline{\psi} (x + a\widehat{\mu}) \, \Gamma \,
U_\mu^\dagger (x) \, \psi (x) \Big) .
\end{equation}
There is also one counterterm of dimension three, which is constructed
from $\Gamma$, i.e.
\begin{equation}
\frac{ic_3 (g_0)}{a} \,\overline{\psi}(x)\,\Gamma\,\psi(x) , 
\end{equation}
which is already present in the bare Bori\c{c}i-Creutz action, albeit
with fixed coefficient $-2/a$. In the general renormalized action the
coefficient of this operator must be tuned. It is convenient in our
perturbative work to use the convention that the operator
$\overline{\psi} (x)\Gamma\psi (x)$ in the renormalized action has as
a coefficient $(-2+c_3)\,i/a$, although for Monte Carlo simulations
other choices might be more appropriate. Since this piece of the
action determines the location of the poles of the propagators, which
are moved by radiative corrections, a possible renormalization
condition is the requirement that the value of the coefficient must
restore the poles to their original positions at $p_1$ and $p_2$.

For Karsten-Wilczek fermions things work out in a similar way.  Here
we are allowed to construct objects in which Kronecker deltas can
constrain any Lorentz index to be equal to~4. It is easy to see that
the only gauge-invariant counterterm of dimension four that can be
added to the bare action is $\overline{\psi}\,\gamma_4 D_4 \psi$.
A suitable discretization for this operator is
\begin{equation}
d_4 (g_0) \, \frac{1}{2a} \Big( \overline{\psi}(x)
\, \gamma_4 \, U_4 (x) \, \psi (x + a\widehat{4}) 
-\overline{\psi} (x + a\widehat{4}) \, \gamma_4 \,
U_4^\dagger (x) \, \psi (x) \Big) .
\end{equation}
The counterterm of dimension three, i.e.
\begin{equation}
\frac{id_3 (g_0)}{a} \,\overline{\psi} (x) \,\gamma_4
\,\psi (x) ,
\end{equation}
is already contained in the bare Karsten-Wilczek action, where it has
a fixed coefficient of~$3/a$. In the fully renormalized theory we will
denote the coefficient of the term $\overline{\psi} (x)\gamma_4\psi
(x)$ by $(3+d_3)\,i/a$.

In perturbation theory the coefficients multiplying the above
counterterms are functions of the coupling, starting at order
$g_0^2$. They generate new vertices and propagator insertions. At one
loop they give rise to additional contributions to fermion lines, and
these insertions must be taken into account for a consistent one-loop
calculation. From the above expressions the rules for the corrections
to external fermion propagators, which will be needed for the
calculations presented in this paper, can be easily derived. These are
independent of the lattice discretization chosen for the
counterterms. Since the propagator is the inverse of the quadratic
part of the action, they are given in momentum space by
\begin{equation}
-ic_4 (g_0) \,\,\Gamma \,\sum_\nu p_\nu , \quad
 -\frac{ic_3(g_0)}{a}\,\Gamma
\label{eq:bcctlines}
\end{equation}
for Bori\c{c}i-Creutz fermions, and
\begin{equation}
-id_4 (g_0) \,\,\gamma_4 \,p_4 , \quad
 -\frac{id_3(g_0)}{a}\,\gamma_4
\label{eq:kwctlines}
\end{equation}
for Karsten-Wilczek fermions, respectively. We will determine all
their coefficients at one loop in perturbation theory (see Section
\ref{sec:fermcts}), by requiring that the renormalized self-energy
assumes its standard Lorentz-invariant form.

We also need counterterms for the pure gauge part of the actions of
minimally doubled fermions. Although at the bare level the breaking of
hyper-cubic symmetry is generated by the fermionic actions only, it
propagates via the interactions between quarks and gluons also to the
gauge sector in the renormalized theories. One effect is that some of
the terms in the purely gluonic part can renormalize with different
factors, and as a consequence pure gauge counterterms must be added 
to the renormalized actions to correct this imbalance. They are of the
(continuum) $\Tr\, FF$ form, but with non-conventional choices of the
indices which reflect the breaking of the Lorentz (hyper-cubic)
symmetry.

Let us consider first the Bori\c{c}i-Creutz case. If we choose all
four indices of $\Tr\, FF$ to appear only once in each summation (as
allowed by hyper-cubic symmetry breaking), we can construct a
counterterm which has the continuum form
\begin{equation}
\sum_{\lambda,\rho,\sigma,\tau} \Tr\, F_{\lambda\rho}(x) \,
F_{\sigma\tau}(x).
\end{equation}
However, since $F_{\mu\nu}$ is antisymmetric in the indices, this is
identically zero. The next possibility is to contract one Lorentz
index shared by both field tensors (as in the usual Einstein
convention), whilst summing individually over the two remaining
indices, i.e.
\begin{equation}
c_{\rm{P}}(g_0) \, \sum_{\lambda,\rho,\tau} \Tr\, F_{\lambda\rho}(x) \,
F_{\rho\tau}(x) . 
\label{eq:ctbcplaq}
\end{equation}
This operator, whose form is reminiscent of the energy-momentum
tensor, is the only possible purely gluonic counterterm for this
action.\,\footnote{In fact, choosing two pairs of summed indices will
reproduce the usual Lorentz invariant term $\sum_{\lambda,\rho} \Tr\,
F_{\lambda\rho}(x) \, F_{\lambda\rho}(x)$, already present in the bare action.}
A lattice counterpart for this counterterm can be obtained by employing the 
widely used ``clover'' expression of the $F_{\mu\nu}$ tensor\,\cite{impr:SW}.

At one loop this counterterm contributes only via insertions in gluon
propagators. Denoting the fixed external indices at both ends of these
lines with $\mu$ and $\nu$, all possible lattice discretizations of
this counterterm yield the same Feynman rule in momentum space,
namely\,\footnote{The lowest order in the expansion of
$F_{\lambda\rho}(x) \, F_{\rho\tau}(x)$ in momentum space is
\begin{equation}
- p_\lambda p_\rho A_\rho A_\tau - p_\tau p_\rho A_\rho A_\lambda 
+ p^2_\rho A_\lambda A_\tau + p_\lambda p_\tau A_\rho A_\rho .
\end{equation}
When the external indices of the gluon propagator are set equal to 
$\mu$ and $\nu$, this becomes
\begin{equation}
- p_\mu p_\lambda A_\mu A_\nu - p_\nu p_\lambda A_\mu A_\nu 
+ p^2 A_\mu A_\nu + p_\lambda p_\tau A_\mu A_\nu \,\delta_{\mu\nu} .
\end{equation}
Here all indices but $\mu$ and $\nu$ are summed.
A similar derivation holds for the Karsten-Wilczek action, where it gives
eq.\, (\ref{eq:kwvpextra}).}
\begin{equation}
-c_{\rm{P}}(g_0) \,\left[ (p_\mu + p_\nu)\,\sum_\lambda p_\lambda
-p^2  - \delta_{\mu\nu}\Big( \sum_\lambda p_\lambda \Big)^2 \right].
\label{eq:ctbcplaqmom}
\end{equation}
As we will see explicitly in Section \ref{sec:vacpol}, the presence of
this counterterm is essential in order to ensure the correct
renormalization of the vacuum polarization.

It is not hard to infer that in the case of Karsten-Wilczek fermions
the temporal plaquettes (the chromo-electric field) renormalize 
differently compared with the spatial plaquettes (corresponding to the
chromo-magnetic field). The counterterm to be introduced will contain
an asymmetry between these two kinds of plaquettes, and can be written
in continuum form as
\begin{equation}
d_{\rm{P}}(g_0) \, 
\sum_{\rho,\lambda} \Tr\, F_{\rho\lambda}(x) \, F_{\rho\lambda}(x) \,
\delta_{\rho 4} . 
\label{eq:ctkwplaq}
\end{equation}
This is the only required gluonic counterterm for this action,
since introducing another factor of $\delta_{\lambda 4}$ in the above
expression will produce a vanishing object. It is immediate to write
down a lattice representation for it, using the plaquette:
\begin{equation}
d_{\rm{P}}(g_0) \,\frac{\beta}{2}\,
\sum_{\rho,\lambda} \left( 1-\frac{1}{N_{\rm{C}}}\,\Tr\,P_{4
  \lambda}(x) \right).
\end{equation}
The Feynman rule for an external gluon line that this counterterm
generates reads
\begin{equation}
- d_{\rm{P}}(g_0) \, \left[ p_\mu p_\nu \,(\delta_{\mu 4 } +
 \delta_{\nu 4 })  -\delta_{\mu\nu} \left( p^2\,\delta_{\mu 4 }
 \delta_{\nu 4 } +p_4^2 \right)  
\right] .
\label{eq:kwvpextra}
\end{equation}
This term guarantees the correct renormalization of the vacuum
polarization.

From the gauge-invariant expressions of the operators introduced in
this section we can read off that interaction vertices are generated
by the counterterms. However, these vertex insertions are of higher
order in $g_0$ (at least of $\rmO(g_0^3)$), and thus they cannot
contribute to the one-loop amplitudes presented in this paper. 

It is worth stressing that the form of the counterterms that we have
constructed remains the same at all orders of perturbation
theory. Only the values of the coefficients change according to the
loop order considered. The same counterterms appear also at the
non-perturbative level, and will be required for a consistent
numerical simulation of these fermions, as we will see in Section
\ref{sec:simul}. Finally, we want to emphasize here that counterterms not
only provide additional Feynman rules for the calculation of loop
amplitudes, but can also modify Ward identities and hence, in
particular, contribute additional terms to the conserved currents, as
we will see explicitly in Section \ref{sec:bilin_cons}.

\section{Fermionic counterterms at one loop}
\label{sec:fermcts}

We now show how the coefficients of the counterterms that appear in
the quark action can be fixed by computing the quark self-energy.
Technical details in the case of Karsten-Wilczek fermions are deferred
to appendix~\ref{app:self}, while the case of Bori\c{c}i-Creutz action
was described at length in \cite{mind:Capitani09}.

In the following we will work in some general covariant gauge, where
$\partial_\mu A_\mu=0$ and $\alpha$ denotes the gauge parameter.
For Bori\c{c}i-Creutz fermions, the result for the one-loop diagrams
of the quark self-energy, without including the counterterms, reads
(see eq.\,(15) of \cite{mind:Capitani09})
\begin{equation}
\Sigma (p,m_0) = i\slash{p}\,\Sigma_1(p) +m_0\,\Sigma_2(p) 
+ c_1 (g_0)\cdot i\, \Gamma \sum_\mu p_\mu
+ c_2 (g_0)\cdot i\, \frac{\Gamma}{a},
\label{eq:totalself}
\end{equation}
where
\begin{eqnarray}
\Sigma_1(p) &=& \frac{g_0^2}{16\pi^2} \,\CF \,\Bigg[ \log a^2p^2
  +6.80663 +(1-\alpha) \Big(-\log a^2p^2 + 4.792010 \Big) \Bigg] ,
\label{eq:Sigma1self} \\ 
\Sigma_2(p) &=& \frac{g_0^2}{16\pi^2} \,\CF \,\Bigg[ 4\,\log
  a^2p^2 -29.48729 +(1-\alpha) \Big(-\log a^2p^2 +5.792010 \Big) \Bigg] ,
\label{eq:Sigma2self} \\ 
c_1 (g_0)\, &=&  1.52766 \cdot\frac{g_0^2}{16\pi^2} \,\CF ,
 \label{eq:c1self} \\
c_2 (g_0)\, &=& 29.54170 \cdot\frac{g_0^2}{16\pi^2} \,\CF ,
\end{eqnarray}
with $C_F=(N_c^2-1)/2N_c$.
From eq.\,(\ref{eq:totalself}) alone, the inverse propagator at
one loop level would assume the form
\begin{equation}
\Sigma^{-1} (p,m_0) = \Big( 1 -\Sigma_1 \Big) \cdot
\Big\{ i\slash{p}
+ m_0 \,\Big( 1 -\Sigma_2 +\Sigma_1 \Big) 
- ic_1 (g_0)\,\Gamma \,\sum_\mu p_\mu
-\frac{ic_2 (g_0)}{a}\,\Gamma \Big\}.
\label{eq:invprop}
\end{equation}
We now show how the coefficients multiplying the counterterms of the
Bori\c{c}i-Creutz quark action can be fixed by imposing the condition
that the renormalized propagator take the standard form
\begin{equation}
\Sigma (p,m_0) = \frac{Z_2}{i\slash{p} + Z_m\, m_0} ,
\label{eq:stdprop}
\end{equation}
where the wave-function and quark mass renormalization factors are
given by
\begin{eqnarray}
Z_2 &=& \Big( 1 -\Sigma_1 \Big)^{-1} ,        \label{eq:z2}  \\
\zm &=& 1 - \Big( \Sigma_2 -\Sigma_1 \Big) ,  \label{eq:zm}
\end{eqnarray}
Indeed, by looking at eq.\,(\ref{eq:bcctlines}) we see that the piece
in eq.\,(\ref{eq:invprop}) which is proportional to $c_1(g_0)$ can
be eliminated by tuning the coefficient of the counterterm of
dimension four, $c_4(g_0)\, \overline{\psi} \Gamma \sum_\mu D_\mu
\psi$, while the one proportional to $c_2(g_0)$ can be eliminated by
a suitable choice of the counterterm of dimension three,
$ic_3(g_0)\,\overline{\psi} \Gamma \psi /a$. After inserting these
counterterms, one has
\begin{eqnarray}
\frac{1}{i\slash{p}+m_0} &+& \frac{1}{i\slash{p}+m_0} \cdot
\Big[ i\slash{p}\,\Sigma_1 +m_0\,\Sigma_2 
+ c_1 \cdot i\, \Gamma \sum_\mu p_\mu + c_2 \cdot i\, \frac{\Gamma}{a}
- c_4 \cdot i\, \Gamma \sum_\mu p_\mu \nonumber \\
& & \phantom{\frac{1}{i\slash{p}+m_0} \cdot \;}
 - c_3 \cdot i\, \frac{\Gamma}{a} \Big]
\cdot \frac{1}{i\slash{p}+m_0} \nonumber \\
& = & \frac{1}{i\slash{p}\,(1-\Sigma_1) +m_0\,(1-\Sigma_2) 
- (c_1 -c_4 ) \cdot i\, \Gamma \sum_\mu p_\mu
- (c_2 -c_3 ) \cdot i\, \frac{\Gamma}{a}} .
\end{eqnarray}
It follows that $c_4 (g_0) = c_1 (g_0)$ and $c_3 (g_0) = c_2
(g_0)$, if eqs.\,(\ref{eq:stdprop})--(\ref{eq:zm}) are to be
recovered. Thus, at this order the coefficients of these counterterms
for Bori\c{c}i-Creutz fermions are determined as
\begin{eqnarray}
c_3 (g_0) &=& 29.54170 \cdot\frac{g_0^2}{16\pi^2}\,\CF+\rmO(g_0^4), \\
c_4 (g_0) &=&  1.52766 \cdot\frac{g_0^2}{16\pi^2}\,\CF+\rmO(g_0^4).
\end{eqnarray}
Of course, in order to be useful for numerical simulations, these
coefficients will have to be determined non-perturbatively, by
imposing suitable renormalization conditions. We will return to this
question in Section \ref{sec:simul}.

The explicit calculation of the self-energy for Karsten-Wilczek
fermions at one loop proceeds along the same lines as for
Bori\c{c}i-Creutz. Further details as well as the individual results
for the tadpole and sunset diagrams can be found in
Appendix\,\ref{app:self}. The result for the total contribution,
without counterterms, is
\begin{equation}
\Sigma (p,m_0) = i\slash{p}\,\Sigma_1(p) +m_0\,\Sigma_2(p) 
+ d_1 (g_0)\cdot i\, \gamma_4 p_4
+ d_2 (g_0)\cdot i\, \frac{\gamma_4}{a},
\label{eq:totalself2}
\end{equation}
where
\begin{eqnarray}
\Sigma_1(p) &=& \frac{g_0^2}{16\pi^2} \,\CF \,\Bigg[ \log a^2p^2
  +9.24089 +(1-\alpha) \Big(-\log a^2p^2 + 4.792010 \Big) \Bigg] ,
\label{eq:Sigma1self2} \\ 
\Sigma_2(p) &=& \frac{g_0^2}{16\pi^2} \,\CF \,\Bigg[ 4\,\log
  a^2p^2 -24.36875 +(1-\alpha) \Big(-\log a^2p^2 +5.792010 \Big) \Bigg] ,
\label{eq:Sigma2self2} \\ 
d_1 (g_0)\, &=&  -0.12554 \cdot\frac{g_0^2}{16\pi^2} \,\CF ,
 \label{eq:c1self2} \\
d_2 (g_0)\, &=& -29.53230 \cdot\frac{g_0^2}{16\pi^2} \,\CF .
\end{eqnarray}
Without any counterterms, the inverse propagator at one loop is
\begin{equation}
\Sigma^{-1} (p,m_0) = \Big( 1 -\Sigma_1 \Big) \cdot
\Big( i\slash{p} + m_0 \,\Big( 1 -\Sigma_2 +\Sigma_1 \Big)
- id_1 \,\gamma_4 p_4 - \frac{id_2}{a} \,\gamma_4 \Big).
\end{equation}
As in the case of Bori\c{c}i-Creutz fermions, the extra contributions
proportional to $d_1(g_0)$ and $d_2(g_0)$ can be cancelled by suitably
tuning the coefficients of the counterterms $d_4(g_0)\,\overline{\psi}
\gamma_4 D_4 \psi$ and $id_3(g_0)\,\overline{\psi} \gamma_4 \psi /a$
respectively. Thus, the full inverse propagator at one loop for
Karsten-Wilczek fermions can be written in the standard form,
eqs.\,(\ref{eq:stdprop})--(\ref{eq:zm}). The coefficients of the
counterterms for Karsten-Wilczek fermions so determined are at
one-loop order
\begin{eqnarray}
d_3 (g_0)&=& -29.53230\cdot\frac{g_0^2}{16\pi^2}\,\CF+\rmO(g_0^4), \\
d_4 (g_0)&=&  -0.12554\cdot\frac{g_0^2}{16\pi^2}\,\CF+\rmO(g_0^4).
\end{eqnarray}
One may expect that the above subtraction procedure can be carried out
consistently at every order of perturbation theory. After the
subtractions via the appropriate counterterms are properly taken into
account, the extra terms appearing in the self-energy can be
eliminated.

Recalling that the term proportional to $\gamma_4$ in the
Karsten-Wilczek action can be multiplied by a parameter $\lambda$, one
may wonder whether a suitable choice of $\lambda$ could eliminate the
power-divergent contribution to the quark self-energy, without
resorting to counterterms. This could in principle be accomplished in
view of the fact that the tadpole and sunset diagrams contribute with
opposite sign to the self-energy, while the tadpole contribution is
linearly proportional to $\lambda$. We have investigated this issue
using our perturbative expressions, and concluded, however, that such
a cancellation cannot take place. The reason is that while the value
of the tadpole decreases as $\lambda$ is lowered, the contribution of
the sunset diagram also decreases at the same time. Since the latter
diagram always remains much smaller than the tadpole it cannot
compensate its value. Thus, it is not possible to eliminate this
power-divergent extra term without counterterms.

\section{Quark bilinears and conserved currents}
\label{sec:bilin_cons}

For many applications, knowledge of the renormalization factors for
quark bilinears is required. These are obtained by computing the
appropriate vertex diagrams and adding the contribution $\Sigma_1$ to
the quark self-energy. For Bori\c{c}i-Creutz fermions, the results for
the vertex diagrams of the various bilinears are given in
\cite{mind:Capitani09}. For Karsten-Wilczek fermions the corresponding
results are listed in Appendix\,\ref{app:bilin}. Here we note that for
the latter, hyper-cubic symmetry breaking induces different radiative
corrections for temporal and spatial components of the vector and
axial-vector currents.

In the following we discuss the issue of the renormalization of the
quark mass. For both realizations of minimally doubled fermions,
chiral symmetry is preserved and so the quark mass does not undergo any
additive renormalization. Thus, the relation between the bare and
renormalized quark masses is
\begin{equation}
   m_{\rm R} = \zm\,m_0,
\end{equation}
where $\zm$ is given in eq.\,(\ref{eq:zm}). The full expression for
the renormalization factors of the scalar and pseudo-scalar densities
in perturbation theory at one loop is
\begin{equation}
   \zs = \zp = 1 - \Big( \Lambda_{\rm S}+\Sigma_1 \Big),
\end{equation}
where $\Lambda_{\rm S}$ is the vertex correction of the scalar
density, which is given by \eq{eq:lambda-s} for Karsten-Wilczek and
eq.\,(25) of \cite{mind:Capitani09} for Bori\c{c}i-Creutz fermions,
respectively. This number is exactly equal to the
$\rmO(g_0^2)$-contribution to $\Sigma_2$ (eq.\,(\ref{eq:Sigma2self})
or (\ref{eq:Sigma2self2})), but comes with an opposite sign:
$\Lambda_{\rm{S}}=-\Sigma_2$. Thus, when we compare them with
eq.\,(\ref{eq:zm}), we see that the renormalization factors $\zs$ and
$\zp$ of the scalar and pseudo-scalar densities satisfy
\begin{equation}
    1/\zm=\zs=\zp,
\end{equation}
where the last equality is a consequence of chiral symmetry. We have
thus verified at one loop that the renormalization of the quark mass
for both variants of minimally doubled fermions presented here has the
same form as, say, in the case of overlap or staggered fermions.

Using the expressions for the vertex corrections of the local vector
and axial-vector currents and taking the renormalization of the
wave-function into account, the corresponding renormalization factors
$\zv$ and $\za$ are not equal to one. In order to construct the
conserved currents, which are protected against renormalization, one
has to derive the chiral Ward identities, for instance, along the
lines of ref.\,\cite{Boch}. To do this, we have to use the expressions
for the lattice actions of Bori\c{c}i-Creutz and Karsten-Wilczek
fermions in position space, which are given in
eqs.\,(\ref{eq:creutz-action-pos}) and (\ref{eq:wilczek-action-pos})
respectively.

It is important to note that the counterterms which have been added to
the actions can also contribute to the chiral Ward identities, and
thus provide new terms in the expressions of the conserved currents.
Actually, it is easy to see that the counterterms of dimension three
do not modify the Ward identities, but the counterterms of dimension
four do. The latter generate additional terms in the Ward identities
and hence also in the conserved currents. Thus, the form of the
conserved currents turns out to be different from what the bare
actions would have given.

It is worth stressing again that, as we already discussed
in\,\cite{mind:Capitani09}, these actions in the massless case are
invariant under a chiral $\rm U(1)\otimes U(1)$ transformation.  The
chiral Ward identities associated with these exact symmetries yield the
currents of the theory. The chiral $\rm U(1)\otimes U(1)$ symmetry 
of minimally doubled fermions implies that as the quark mass 
is tuned to zero there is only one Goldstone boson, which can be
naturally considered to be the neutral pion. The charged pions will 
instead be massive in the chiral limit (at non-zero lattice spacing). 

If one applies the standard vector and axial transformations, i.e.
\begin{eqnarray}
& &\delta_V \psi = i \alpha_V \psi , \qquad\phantom{\gamma_5}
   \delta_V \overline{\psi} = - i \overline{\psi} \alpha_V , \nonumber \\
& &\delta_A \psi = i \alpha_A \gamma_5 \psi , \qquad
   \delta_A \overline{\psi} = i \overline{\psi} \alpha_A \gamma_5 ,
\end{eqnarray}
under which the Lagrangian remains invariant, one can identify the
conserved vector current for Bori\c{c}i-Creutz fermions in the
renormalized theory as
\begin{eqnarray}
V_\mu^{\mathrm c} (x) &=& \frac{1}{2} \bigg(
   \overline{\psi} (x) \, (\gamma_\mu+i\,\gamma'_\mu) \, U_\mu (x) \,
   \psi (x+a\widehat{\mu}) 
 + \overline{\psi} (x+a\widehat{\mu}) \, (\gamma_\mu-i\,\gamma'_\mu) \,
   U_\mu^\dagger (x) \, \psi (x) \bigg) \nonumber \\
 && +\frac{c_4 (g_0)}{2} \,
 \bigg( \overline{\psi} (x) \, \Gamma \, 
   U_\mu (x) \, \psi (x+a\widehat{\mu}) 
 + \overline{\psi} (x+a\widehat{\mu}) \, \Gamma \, 
   U_\mu^\dagger (x) \, \psi (x) \bigg) .  
\label{eq:noether-vector}
\end{eqnarray}
The axial-vector current (which is only conserved in the massless case) 
is given by
\begin{eqnarray}
A_\mu^{\mathrm c} (x) &=& \frac{1}{2} \bigg(
   \overline{\psi} (x) \, (\gamma_\mu+i\,\gamma'_\mu) \gamma_5 \,
   U_\mu (x) \,
   \psi (x+a\widehat{\mu})
  + \overline{\psi} (x+a\widehat{\mu}) \, (\gamma_\mu-i\,\gamma'_\mu)
   \gamma_5
   \, U_\mu^\dagger (x) \, \psi (x) \bigg) \nonumber \\
 && +\frac{c_4 (g_0)}{2} \,
 \bigg( \overline{\psi} (x) \, \Gamma \gamma_5 \, 
   U_\mu (x) \, \psi (x+a\widehat{\mu})  
  + \overline{\psi} (x+a\widehat{\mu}) \, \Gamma \gamma_5 \, 
   U_\mu^\dagger (x) \, \psi (x) \bigg) .  
\label{eq:noether-axial}
\end{eqnarray}
For Karsten-Wilczek fermions the conserved vector current turns out to be
\begin{eqnarray}
V_\mu^{\mathrm c} (x) &=& \frac{1}{2} \bigg(
   \overline{\psi} (x) \, (\gamma_\mu -i\gamma_4 \, (1-\delta_{\mu 4}) ) \, 
   U_\mu (x) \, \psi (x+a\widehat{\mu}) \nonumber \\ 
 && \qquad \qquad + \overline{\psi} (x+a\widehat{\mu}) \, 
   (\gamma_\mu +i\gamma_4 \, (1-\delta_{\mu 4}) )  \, 
   U_\mu^\dagger (x) \, \psi (x) \bigg) \nonumber \\
 && + \frac{d_4 (g_0)}{2} \bigg(
   \overline{\psi} (x) \, \gamma_4 \, U_4 (x) \, \psi (x+a\widehat{4})  
  + \overline{\psi} (x+a\widehat{4}) \, \gamma_4 \, 
   U_4^\dagger (x) \, \psi (x) \bigg) ,  
\label{eq:noether-vector2}
\end{eqnarray}
and the axial-vector current is given by
\begin{eqnarray}
A_\mu^{\mathrm c} (x) &=& \frac{1}{2} \bigg(
   \overline{\psi} (x) \, (\gamma_\mu -i\gamma_4 \, (1-\delta_{\mu 4}) ) 
   \gamma_5 \, U_\mu (x) \, 
   \psi (x+a\widehat{\mu}) \nonumber \\ 
 && \qquad \qquad + \overline{\psi} (x+a\widehat{\mu}) \, 
    (\gamma_\mu +i\gamma_4 \, (1-\delta_{\mu 4}) ) \gamma_5 
   \, U_\mu^\dagger (x) \, \psi (x) \bigg) \nonumber \\
 && + \frac{d_4 (g_0)}{2} \bigg(
   \overline{\psi} (x) \, \gamma_4 \gamma_5 \, U_4 (x) \, 
   \psi (x+a\widehat{4}) 
  + \overline{\psi} (x+a\widehat{4}) \, \gamma_4 \gamma_5 
   \, U_4^\dagger (x) \, \psi (x) \bigg) . 
\label{eq:noether-axial2}
\end{eqnarray}
It is instructive to verify that the renormalization constants of
these currents is equal to one, since this exercise illustrates the
important r\^ole of the counterterms. Here we restrict a detailed
discussion to the conserved vector current for Karsten-Wilczek
fermions, noting that this is entirely analogous to the
Bori\c{c}i-Creutz case. Furthermore, the corresponding expressions for
the conserved axial-vector current in both discretizations are
trivially obtained from the formulae below by replacing $\gamma_\mu$
by $\gamma_\mu\gamma_5$ and $\gamma_4$ with $\gamma_4\gamma_5$.

The renormalization factor of the vector current is given by
\begin{equation}
\zv=1-(\Lambda_{\rm{V}}+\Sigma_1),
\end{equation}
where $\Lambda_{\rm{V}}$ denotes the vertex correction and $\Sigma_1$
the self-energy. The sum of the vertex (diagram~(a) in
Fig.\,\ref{fig:diagrams}), the ``sails'' (diagrams (b) and (c)) and
the operator tadpole (diagram (d)) corresponds to the first two lines
in \eq{eq:noether-vector2} and yields
\begin{equation}
\frac{g_0^2}{16\pi^2} \,\CF \,\gamma_\mu\,\bigg[ -\log a^2p^2 -9.24089 
+\delta_{\mu 4}\cdot 0.12554 
+(1-\alpha) \Big(\log a^2p^2 -4.79201 \Big) \bigg] .
\label{eq:vert_tad_sails}
\end{equation}
Now, a counterterm is also included in the expression
of the conserved current (c.f. the last line of
eq.\,(\ref{eq:noether-vector2})), and contributes a factor of
\begin{equation}
d_4(g_0) \,\gamma_4  = d_4(g_0) \,\gamma_\mu \delta_{\mu 4}
\end{equation}
to its renormalization factor. At lowest order in perturbation theory
this was evaluated as part of the determination of the self-energy,
with the result
\begin{equation}
d_4(g_0) = - 0.12554 \cdot\frac{g_0^2}{16\pi^2} \,\CF +\rmO(g_0^4).
\end{equation}
This cancels exactly the contribution proportional to $\delta_{\mu 4}$
in \eq{eq:vert_tad_sails}. So, apart from including the wave-function
renormalization, the result for the proper diagrams is
\begin{equation}
\Lambda_{\rm{V}}=
\frac{g_0^2}{16\pi^2} \,\CF \,\gamma_\mu\,\bigg[ -\log a^2p^2 -9.24089 
+(1-\alpha) \Big(\log a^2p^2 -4.79201 \Big) \bigg] .
\end{equation}
Finally, we see that this expression is equal and opposite to the
contribution of $\Sigma_1(p)$ of the quark self-energy,
\eq{eq:Sigma1self2}, and so exactly compensates it. Thus, we conclude
that the renormalization factor of these point-split currents is
one. This holds to all significant digits that we have achieved, and
confirms that the expressions that we have derived via the chiral Ward
identities are indeed conserved currents.

For Bori\c{c}i-Creutz fermions, the sum of vertex, sails and operator
tadpole diagrams for the conserved vector current is
\cite{mind:Capitani09}
\begin{equation}
\frac{g_0^2}{16\pi^2} \,\CF \,\gamma_\mu\,\bigg[ -\log a^2p^2 -6.80664
+(1-\alpha) \Big(\log a^2p^2 -4.79201 \Big) \bigg]
-1.52766 \cdot \frac{g_0^2}{16\pi^2} \,\CF \cdot \Gamma .
\label{eq:bc-vstsum}
\end{equation}
Again, at this point one has to add the contribution from the
counterterm of dimension four in the conserved current (i.e. the last line of
\eq{eq:noether-vector}),
\begin{equation}
c_4(g_0) \, \Gamma ,
\end{equation}
whose coefficient was already fixed by the result of the one-loop
self-energy: 
\begin{equation}
c_4(g_0) = 1.52766 \cdot\frac{g_0^2}{16\pi^2} \,\CF +\rmO(g_0^4).
\end{equation}
This cancels the Lorentz non-invariant term in eq.\,(\ref{eq:bc-vstsum}). 
After including the contribution of the wave-function renormalization 
it can be easily seen that the renormalization constant of this current 
is equal to one as well.

\section{Vacuum polarization}
\label{sec:vacpol}

In order to gain a deeper understanding of the properties of minimally 
doubled fermions, we have also taken the task to calculate the one-loop 
vacuum polarization of the gluon for our two realizations of minimally
doubled fermions.

Here we focus on the radiative corrections to the bare gluon
propagator which arise from fermion loops and the gluonic counterterm.
At one loop the perturbative contributions to the vacuum polarization 
due to loops of gluons and ghosts are independent of the chosen 
fermionic lattice action, and are thus irrelevant for the problem 
we are studying. However, quark loops are able to generate
hyper-cubic-breaking terms, and here we show that this is what indeed
happens for both Karsten-Wilczek and Bori\c{c}i-Creutz fermions.

It is instructive to recall the expression for one flavour of Wilson
fermions. In this case, neither breaking of hyper-cubic symmetry nor
fermion doubling take place, and the result is
\begin{equation}
\Pi^{(f)}_{\mu\nu} (p) = \Bigg( p_\mu p_\nu-\delta_{\mu\nu}p^2 \Bigg) \,
\Bigg[\frac{g_0^2}{16\pi^2} C_2 \Bigg( -\frac{4}{3} \log p^2a^2 +4.337002
\Bigg) \Bigg] ,
\end{equation}
where $C_2$ is defined via $\Tr \,(t^at^b) = C_2 \,\delta^{ab}$. It is
straightforward to see that this gauge invariant result satisfies the
Ward identity $p^\mu \Pi^{(f)}_{\mu\nu} (p)=0$, which expresses the
conservation of the fermionic current.

The result of our calculations for the above quantity using
Bori\c{c}i-Creutz fermions, without including the gluonic counterterm,
reads
\begin{eqnarray}
\Pi^{(f)}_{\mu\nu} (p) &=& \Bigg( p_\mu p_\nu-\delta_{\mu\nu}p^2
\Bigg)  
\Bigg[\frac{g_0^2}{16\pi^2} C_2 \Bigg( -\frac{8}{3} \log p^2a^2
 +23.6793 \Bigg) \Bigg] \\
&& 
- \Bigg( (p_\mu + p_\nu)\,\sum_\lambda p_\lambda -p^2 
- \delta_{\mu\nu}\Big( \sum_\lambda p_\lambda \Big)^2 \Bigg) \,
\frac{g_0^2}{16\pi^2} C_2 \cdot 0.9094
\nonumber .
\end{eqnarray}
By comparing the coefficient of the divergence to the expression for
Wilson fermions we see explicitly that this result corresponds to two
flavours of quarks: Each of the two doublers contributes an equal
amount to the divergence. The same is true for Karsten-Wilczek
fermions, for which our calculation gives the result (again without
counterterms)
\begin{eqnarray}
\Pi^{(f)}_{\mu\nu} (p) &=& \Bigg( p_\mu p_\nu-\delta_{\mu\nu}p^2
\Bigg)  
\Bigg[\frac{g_0^2}{16\pi^2} C_2 \Bigg( -\frac{8}{3} \log p^2a^2 
 +19.99468 \Bigg) \Bigg] \\
&& - \Bigg( p_\mu p_\nu \,(\delta_{\mu 4 } + \delta_{\nu 4 }) 
-\delta_{\mu\nu} \left( p^2\,\delta_{\mu 4 } \delta_{\nu 4 } +p_4^2
\right) \Bigg)\, 
\frac{g_0^2}{16\pi^2} C_2 \cdot 12.69766 
\nonumber  .
\end{eqnarray}
We first note, looking at the second lines of the above equations,
that additional terms appear, compared with a standard situation like
Wilson fermions. A remarkable observation is the fact that although
each of the two actions breaks hyper-cubic symmetry, the extra terms
still satisfy the Ward identity $p^\mu \Pi^{(f)}_{\mu\nu} (p)=0$,
expressing current conservation, as can be easily verified. 
It is easy to see that one cannot obtain smaller expressions 
for the symmetry-breaking terms which are symmetric in $\mu$ and $\nu$ 
and still conserve the current. Moreover, our one loop results 
are the most general symmetric quadratic functions of $p$ which 
can be constructed using the hyper-cubic breaking objects 
$\delta_{\mu 4}$ or $\sum_\lambda p_\lambda$.

The correct renormalization of the polarization of the vacuum requires
the inclusion of the counterterm of the pure gauge action. To see how
this works in detail, we consider for illustration Bori\c{c}i-Creutz
fermions. Adding the one-loop results to the tree-level expression,
and exploiting $p^\mu \Pi^{(f)}_{\mu\nu} (p)=0$ in the usual way, we can
write
\begin{eqnarray}
\frac{\delta_{\mu\nu}-\alpha\,\frac{p_\mu p_\nu}{p^2}}{p^2} &+& 
\frac{\delta_{\mu\lambda}-\alpha\,\frac{p_\mu p_\lambda}{p^2}}{p^2} 
\cdot \Big[ (p_\lambda p_\rho-\delta_{\lambda\rho}p^2) \, \Pi(p^2)
  \nonumber \\ 
&& + \Big((p_\lambda + p_\rho)\,\sum_\tau p_\tau -p^2 
- \delta_{\lambda\rho}\Big( \sum_\tau p_\tau \Big)^2 \Big) \, \widetilde{\Pi}(p^2)
\Big] 
\cdot 
\frac{\delta_{\rho\nu}-\alpha\,\frac{p_\rho p_\nu}{p^2}}{p^2} 
\nonumber \\
& = & \frac{\delta_{\mu\nu}- \widetilde{\alpha}\, \frac{p_\mu
    p_\nu}{p^2}}{p^2(1-\Pi(p^2))} \, + \, \frac{(p_\mu +
  p_\nu)\,\sum_\tau p_\tau -p^2 
- \delta_{\mu\nu}\Big( \sum_\tau p_\tau \Big)^2}{p^4} \, \widetilde{\Pi}(p^2) .
\end{eqnarray}
The function $\Pi(p^2)$ in the first line of the above equation multiplies
the standard Lorentz invariant one-loop expression, which is then 
rearranged such as to produce the functional form of the continuum 
tree-level gluon propagator in the last line. From this one can read off 
that the gauge parameter is renormalized according to 
$\widetilde{\alpha} = \alpha\,(1-\Pi)+\Pi$, and that $Z_3 = 1/(1-\Pi(0))$. 
The remaining terms, the ones proportional to $\widetilde{\Pi}(p^2)$, 
are those that break hyper-cubic symmetry, and they cannot be rearranged 
in a similar way.

It is thus evident that these hyper-cubic-breaking contributions must be
eliminated, and this can be achieved by employing the gluonic
counterterms which we introduced in
Section\,\ref{sec:counterterms}. Indeed, the expression for the
gluonic counterterm for Bori\c{c}i-Creutz fermions in momentum space,
\eq{eq:ctbcplaqmom}, is structurally identical to the additional terms
in the vacuum polarization. Requiring the one-loop vacuum polarization
to assume the standard Lorentz invariant form (that is, only the first
term in the above result) then uniquely determines the coefficient of
the counterterm. The non-standard contributions are cancelled for
Bori\c{c}i-Creutz fermions if we set
\begin{equation}
c_{\rm{P}}(g_0)= -0.9094 \cdot\frac{g_0^2}{16\pi^2}\,C_2 +\rmO(g_0^4).
\end{equation}
The reasoning for Karsten-Wilczek fermions is entirely analogous, and
one concludes that
\begin{equation}
d_{\rm{P}}(g_0)= -12.69766 \cdot\frac{g_0^2}{16\pi^2}\,C_2 +\rmO(g_0^4).
\end{equation}
The most important thing to realize is that there are no
power-divergences in our results for the vacuum polarization. In
principle such divergences could arise with coefficients proportional
to $1/a^2$ or $1/a$. We have explicitly checked in our calculations
that the $1/a^2$ tadpole contribution, when non-zero, is in all cases
of equal magnitude and opposite sign with respect to the sunset
diagram.\,\footnote{It is interesting to note that the numbers 
for these diagrams are much larger than in the case of Wilson fermions, 
where the coefficient of $g_0^2 C_2/16\pi^2$ for the tadpole is $-9.67590$. 
For Karsten-Wilczek fermions this number turns out to be $-36.31464$ 
for each spatial component and $7.12931$ for the temporal component.
For Bori\c{c}i-Creutz fermions it is even larger, $-73.71980$.}

We can understand on general grounds why such power-divergences cannot 
appear. To construct hyper-cubic breaking terms one has to employ objects 
like $\Gamma$ and $\sum_\mu p_\mu$ (for Bori\c{c}i-Creutz fermions) and 
$\gamma_4$ and $p_4$ (for Karsten-Wilczek fermions). 
However, after the traces of the fermions loops are evaluated 
no Dirac structures are left over, and momenta cannot anyway appear 
at the $1/a^2$ level. Linear pieces in the momenta, which would be 
required in case of a $1/a$ power divergence, are instead prohibited 
by the symmetry of the diagrams. 

We have also discovered that the hyper-cubic-breaking terms can be put 
for both actions in the same algebraic form, namely
\begin{equation}
p^2 \{\gamma_\mu,\Gamma\} \{\gamma_\nu,\Gamma\}
+ \delta_{\mu\nu} \{\slash{p},\Gamma\}\{\slash{p},\Gamma\}
-\frac{1}{2}\,\{\slash{p},\Gamma\} 
\Big( \{\gamma_\mu,\slash{p}\} \{\gamma_\nu,\Gamma\} 
     +\{\gamma_\nu,\slash{p}\} \{\gamma_\mu,\Gamma\} \Big) ,
\end{equation}
where in the case of Karsten-Wilczek fermions $\Gamma$ must be
replaced by $\gamma_4/2$. This substitution is suggested by comparison
of the standard relation $\Gamma =
\frac{1}{4}\,\sum_{\mu}(\gamma_\mu+\gamma'_\mu)$ of Bori\c{c}i-Creutz
fermions with the formula $\gamma_4 =
\frac{1}{2}\,\sum_{\mu}(\gamma_\mu+\gamma'_\mu)$ for Karsten-Wilczek
fermions, expressing the symmetries of the action (as can be seen from
Section \ref{sec:actions_etc}, when one expands the propagator of the
latter action around the second Fermi point).  Whether there is any
deeper significance to this structural ``equivalence'' of the
hyper-cubic-breaking structures in the vacuum polarizations remains an
open question.

\section{Numerical simulations}
\label{sec:simul}

In this section we discuss the implications of our one-loop
perturbative calculations for numerical simulations of minimally
doubled fermions. The first thing to note is that simulations must be
based on the complete renormalized actions, including the
counterterms. In position space the full expression for the
Bori\c{c}i-Creutz action reads
\begin{eqnarray}
& & S^f_{\rm BC} = a^4 \sum_{x} \left\{ \frac{1}{2a} \sum_{\mu=1}^4
    \Big[\overline{\psi} (x) \, (\gamma_\mu + c_4(g_0) \, \Gamma 
           + i\gamma'_\mu) \, U_\mu (x) \, \psi (x + a\widehat{\mu}) 
\right.\nonumber \\
&& \phantom{S^f = a^4 \sum_{x} \bigg[\frac{1}{2a} \sum_{\mu}\bigg]}
-\overline{\psi} (x + a\widehat{\mu}) \, (\gamma_\mu 
      - c_4(g_0) \, \Gamma - i\gamma'_\mu) \,
   U_\mu^\dagger (x) \, \psi (x) \Big] \nonumber \\
&& \phantom{S^f = a^4 \sum_{x}} 
  + \overline{\psi}(x) \, \Big(m_0+\widetilde{c}_3(g_0)\,
    \frac{i\,\Gamma}{a}\Big) 
    \, \psi (x) \label{eq:BCfullaction} \\
&& \phantom{S^f = a^4 \sum_{x}}
  + \left.\beta \sum_{\mu < \nu} \Bigg( 1 - \frac{1}{N_{\rm{C}}}
    {\rm Re}\,\Tr\,P_{\mu\nu} \Bigg)
  + c_{\rm{P}}(g_0) \, \sum_{\mu,\nu,\rho} \Tr\, 
        \widehat{F}_{\mu\rho}(x) \, \widehat{F}_{\rho\nu}(x) \right\},
\nonumber
\end{eqnarray}
where we have redefined the coefficient of the dimension-three
counterterm, via $\widetilde{c}_3(g_0)=-2+c_3(g_0)$, and $\widehat{F}$
is some lattice discretization of the field-strength tensor. The
renormalized action for Karsten-Wilczek fermions reads
\begin{eqnarray}
& & S^f_{\rm KW} = a^4 \sum_{x} \left\{ \frac{1}{2a} \sum_{\mu=1}^4
    \Big[\overline{\psi}(x)\,(\gamma_\mu(1+ c_4(g_0)\,\delta_{\mu 4})
  -i\gamma_4 \, (1-\delta_{\mu 4}) ) \, U_\mu (x) \, \psi (x +
    a\widehat{\mu}) \right. \nonumber \\ 
&& \phantom{S^f = a^4 \sum_{x} \bigg[ \frac{1}{2a} \sum_{\mu}\bigg]}
-\overline{\psi} (x + a\widehat{\mu}) \, 
(\gamma_\mu(1- d_4(g_0)\,\delta_{\mu 4}) +i\gamma_4 \,
(1-\delta_{\mu 4}) )\,U_\mu^\dagger (x)\, \psi(x)\Big] \nonumber \\
&& \phantom{S^f = a^4 \sum_{x}}
+ \overline{\psi}(x) \, \Big(m_0+\widetilde{d}_3(g_0)\,
  \frac{i\,\gamma_4}{a}\Big) 
  \, \psi (x) \label{eq:KWfullaction} \\
&& \phantom{S^f = a^4 \sum_{x}}\left.
  + \beta \sum_{\mu < \nu} \Bigg( 1 - \frac{1}{N_{\rm{C}}}
   {\rm Re}\,\Tr\,P_{\mu\nu} \Bigg)
    \, \Big( 1 + d_{\rm{P}}(g_0) \, \delta_{\mu 4} \Big)\right\} , \nonumber 
\end{eqnarray}
where $\widetilde{d}_3(g_0)=3+d_3(g_0)$. In perturbation theory the
coefficients of the counterterms admit the expansions
\begin{eqnarray}
\widetilde{c}_3(g_0) & = & -2 + c_3^{(1)} g_0^2 + c_3^{(2)} g_0^4 +
\dots\,, \nonumber \\ 
c_4(g_0) & = & \phantom{-2 +~} c_4^{(1)} g_0^2 + c_4^{(2)} g_0^4 +
\dots\,, \\ 
c_{\rm{P}}(g_0) & = & \phantom{-2 +~} c_{\rm{P}}^{(1)} g_0^2 +
c_{\rm{P}}^{(2)} g_0^4 + \dots\,,  \nonumber
\end{eqnarray}
and 
\begin{eqnarray}
\widetilde{d}_3(g_0) & = & 3 + d_3^{(1)} g_0^2 + d_3^{(2)} g_0^4 +
\dots\,, \nonumber \\ 
d_4(g_0) & = & \phantom{3 +~} d_4^{(1)} g_0^2 + d_4^{(2)} g_0^4 +
\dots\,, \\ 
d_{\rm{P}}(g_0) & = & \phantom{3 +~} d_{\rm{P}}^{(1)} g_0^2 +
d_{\rm{P}}^{(2)} g_0^4 + \dots \,.\nonumber 
\end{eqnarray}
In order to define the full actions at the non-perturbative level one
must impose suitable renormalization conditions which fix the values
of the counterterms beyond perturbation theory. Below we discuss
possible scenarios how this could be achieved.

It is fairly straightforward to determine the coefficients of the
counterterms of dimension four via a non-perturbative condition: as we
have seen in Section\,\ref{sec:bilin_cons}, these counterterms ensure
that the conserved currents have unit renormalization. The
coefficients $c_4$ and $d_4$ can then be fixed by requiring that the
electric charge be equal to one. To this end one can compute suitable
ratios of three-point and two-point correlation functions, involving
the expressions in eqs.\,(\ref{eq:noether-vector})
and~\,(\ref{eq:noether-vector2}), respectively. Adjusting the
coefficients until the charge is unity fixes the values of $c_4$ and
$d_4$. It is an empirical fact that ratios of correlators are obtained
with good statistical accuracy.

Furthermore, radiative corrections induce a shift of the poles of the
quark propagator away from their tree-level positions. Provided that
their coefficients are appropriately tuned, the counterterms of
dimension three ensure that the two Fermi points can be moved back to
their original locations. It is important to realize that radiative
corrections, when moving the poles, do not introduce a sign problem
into the Monte Carlo generation of configurations.  The gauge action
remains real, and the eigenvalues of the Dirac operator are always in
complex conjugate pairs, making the fermion determinant always
non-negative.

On the other hand, these shifts can introduce oscillations as a
function of separation into some hadronic correlation functions.  Such
oscillations, familiar from the staggered formulation, come about
since the underlying fermion field can create several different
species, and these species occur in different regions of the Brillouin
zone. It would be interesting to explore whether or not these 
oscillations could be cancelled by constructing hadronic operators spread 
over nearby neighbours.

It is important to remember that because the two species are of
opposite chirality, the naive $\gamma_5$ matrix is physically a flavour
non-singlet.  The naive on-site pseudoscalar field
$\overline\psi\gamma_5\psi$ can create only flavour non-singlet
pseudoscalar states.  To create the flavour-singlet pseudoscalar meson,
which gets its mass from the anomaly, one needs to combine fields on
nearby sites with appropriate phases.

The purely gluonic counterterm for Bori\c{c}i-Creutz fermions
introduces new operators into the renormalized action, in which
chromo-electric and chromo-magnetic fields enter.  Since the positive
diagonal of space time hypercubes is selected as special, terms
involving $\sum_{\mu,\nu,\rho} F_{\mu\rho} F_{\nu\rho}$ can enter.  This gives 
rise to combinations like $E\cdot B$, $E_1 E_2$, $B_2 B_3$ and similar.
These cross terms can be removed by a diagonalization process,
essentially a rotation redefining the time direction to be along the
positive diagonal.  However, then the coefficents of $E^2$ and $B^2$
can differ.  Effectively, the speed of light for the gluons has been
renormalized.  The coefficient $c_{\rm{P}}$ could then be fixed by
tuning its value to restore the $E$, $B$ symmetry.  The above effects
could turn out to be small, given that at tree level only the
fermionic actions break hyper-cubic symmetry. It could also happen
that other derived quantities are more sensitive to this coefficient
and more suitable for its non-perturbative determination. In general,
one can look for Ward identities and study their deviation from the
standard Lorentz invariant form, as a function of $c_{\rm{P}}$.

For Karsten-Wilczek fermions the purely gluonic counterterm induces an
asymmetry between the plaquettes containing a temporal link relative
to those involving spatial links only. Fixing the coefficient of this
counterterm, $d_{\rm{P}}$, could then be accomplished by computing a
spatial plaquette or Wilson loop, and then equating its result to its
counterpart with components in the time direction.

Eventually Monte Carlo simulations will reveal the actual amount of
symmetry breaking, which could turn out to be large or small,
depending on the observable considered. One important such quantity is
the mass splitting of the charged pions relative to the neutral pion.
Furthermore, the relative magnitude of these symmetry-breaking effects
could be substantially different for Bori\c{c}i-Creutz and
Karsten-Wilczek fermions. Which of these two realizations has the
potential to become the preferred choice for numerical simulations,
will largely depend on this issue.

So far we have not touched on the important subject of identifying the
leading lattice artefacts associated with minimally doubled fermions. 
As we observed in connection with eqs.\,(\ref{eq:bc-covder}) 
and~(\ref{eq:kw-covder}), both realizations of minimally doubled fermions
contain a dimension-five operator in the bare action, and this leads one 
to expect the leading lattice artefacts to be of order~$a$.\,\footnote{This 
contradicts the findings of the (numerical) tree-level analysis of 
ref.\,\cite{Cichy:2008gk}, according to which cutoff effects for pion masses 
and decay constants were of order~$a^2$.}
Naively, one might assume that the preservation of chiral symmetry would 
automatically ensure $\rmO(a)$ improvement, but here we can see that this 
is not always the case. In minimally doubled fermion actions there are in fact 
manifest $\rmO(a)$ effects, which arise as a consequence of the breaking 
of hyper-cubic (and not chiral) symmetry. 

This naturally leads to a discussion of the subject of $\rmO(a)$
improvement.  Since these actions are not improved from the beginning,
at least one further dimension-five operator will be needed in order
to cancel $\rmO(a)$ contributions in on-shell matrix elements. Of
course all possible operators of dimension five that are consistent
with the symmetries must be considered.  Since minimally doubled
fermions respect chiral symmetry, this excludes operators like the
``clover'' term, which contains $\sigma_{\mu\nu}$ and therefore does
not anticommute with $\gamma_5$.  We attempted a cursory analysis, but
there appear to be quite a number of possible operators, even after
some of them are reabsorbed into a rescaling of parameters or
eliminated using the equations of motion.  Just to give an example,
among the possible operators for Bori\c{c}i-Creutz fermions we find
$\overline{\psi} \,\Gamma \sum_{\mu,\nu} D_\mu D_\nu \psi$ and
$\sum_{\mu,\nu,\lambda} F_{\mu\nu} D_\lambda F_{\mu\nu}$.  We
emphasize that such additional dimension-five operators can occur not
only in the quark sector, but also in the pure gauge part. In fact,
when Lorentz invariance is broken, the statement that only operators
with even dimension can appear in the pure gauge action is no longer
true. As this discussion shows, the issue of $\rmO(a)$ improvement can
be quite intricate for this particular type of fermionic
discretizations.  The task of classifying the minimal set of
independent operators is an interesting problem but beyond the scope
of this paper.

We close this section with some speculations.  First, it is not entirely 
clear that we need to move the Fermi points back to the free theory 
positions. As long as there are still two poles in the propagator, 
we have minimal doubling. Further investigations could reveal whether 
relaxing this condition simplifies the tuning of counterterms.
Finally, if we are willing to live with an induced anisotropy, we only
need to tune $c_3$ and $c_4$ roughly and then use physical observables
to measure the remnant anisotropy.  Then only $c_P$ needs to be
adjusted so that the anisotropy for the gluons is the same as for the
fermions.

\section{Conclusions}
\label{sec:concl}

Following on the first investigation of minimally doubled fermions for
Bori\c{c}i-Creutz fermions\,\cite{mind:Capitani09}, we have presented
further perturbative study including another realization, namely
Karsten-Wilczek fermions. At the same time, we have re-interpreted our
analysis of the renormalization of both these actions in the context
of a solid field-theoretical framework.

Our investigations show that both Bori\c{c}i-Creutz and
Karsten-Wilczek fermions are described by a fully coherent quantum
field theory. Since the complete set of operators allowed by the
symmetries of these fermions includes several not present in the
tree-level actions, counterterms must be introduced for a consistent
renormalized theory. After adding the counterterms, only a small
number of new mixings arises for the matrix elements of local
bilinears, none of which is power divergent. For the local vector and
axial-vector currents, some finite mixing with hyper-cubic breaking
operators occurs. For the scalar and pseudoscalar densities and the
tensor operator, on the other hand, the structure of the mixings is
as in the continuum.

We have also constructed the conserved vector and axial-vector
currents for both kinds of fermions.  The vector current is
isospin-singlet, representing the conservation of fermion number.  The
axial current, however, is a non-singlet because the doubled fermions
have opposite chirality.  These currents have simple expressions which
involve only nearest-neighbour points, and do not undergo any
mixing. We have verified at one loop in perturbation theory that their
renormalization constants are equal to one. One of the most attractive
features of Bori\c{c}i-Creutz and Karsten-Wilczek fermions is that
they belong to the very few lattice discretizations that yield a
simple and essentially ultralocal expression for a conserved
axial-vector current. These features could turn out to be a key
advantage in numerical simulations. Furthermore, we have also
calculated the polarization of the vacuum for both actions. We have
proven that, in spite of the breaking of hyper-cubic symmetry, no
power divergences appear for the vacuum polarization of the gluon.

In summary, we have constructed the renormalized theory up to and
including $O(g_0^2)$ for the two realizations of minimally doubled
fermions considered. We have discussed perturbative and
non-perturbative conditions for fixing the coefficients of the three
counterterms required for both realizations.  We have argued that
under reasonable assumptions and following the determination of these
counterterms, no special features of these two realizations of
minimally doubled fermions should hinder their successful Monte Carlo
simulation.

Some questions merit further consideration: firstly, one should
revisit the problem of formulating conditions which allow for a
precise determination of the coefficients of the dimension-three
counterterm at the non-perturbative level, and the same applies to the
gluonic counterterms. Secondly, attempts to improve convergence
towards the continuum limit must take account of the inherent
hyper-cubic symmetry breaking and the induced mixing with
dimension-five operators.

\section*{Acknowledgments}
 
We warmly thank Martin L\"uscher for clarifying discussions and useful
suggestions, especially concerning the counterterms.
This work was supported by Deutsche Forschungsgemeinschaft
(SFB443), the GSI Helmholtz-Zentrum f\"ur Schwerionenforschung, and
the Research Centre ``Elementary Forces and Mathematical Foundations''
(EMG) funded by the State of Rhineland-Palatinate. MC was supported
by contract number DE-AC02-98CH10886 with the U.S.~Department of
Energy.  Accordingly, the U.S. Government retains a non-exclusive,
royalty-free license to publish or reproduce the published form of
this contribution, or allow others to do so, for U.S.~Government
purposes.  MC is particularly grateful to the Alexander von Humboldt
Foundation for support for multiple visits to the University of Mainz.

\begin{appendix}
\section{One-loop calculations for Karsten-Wilczek fermions}

\begin{figure}[t]
\begin{center}
\includegraphics[width=9.cm]{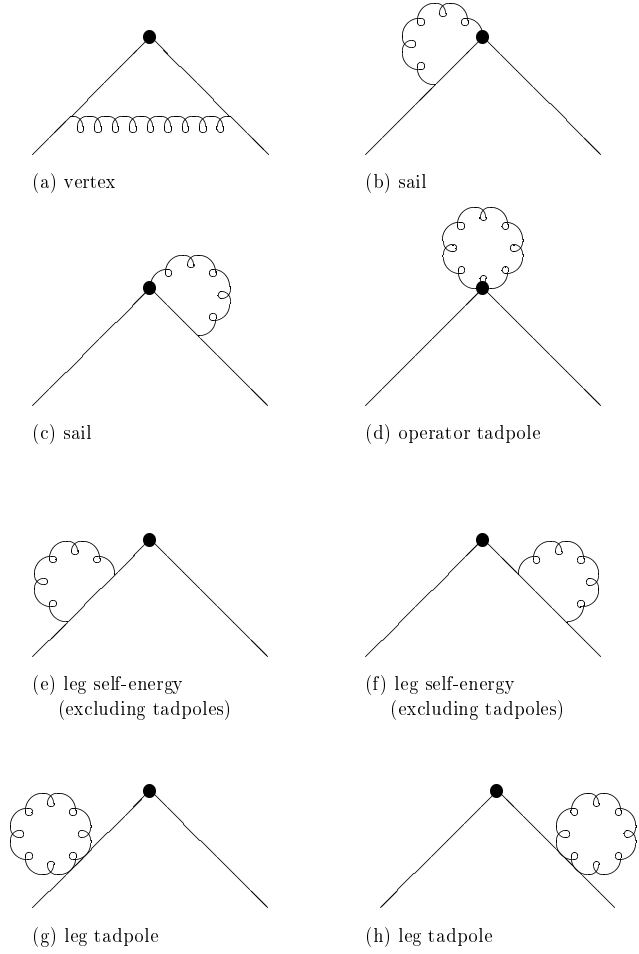}
\caption{\label{fig:diagrams}\small The diagrams needed for the
  one-loop
renormalization of the lattice operators.}
\end{center}
\end{figure}

In this appendix we report the individual expressions at one-loop for
various quantities computed using the Karsten-Wilczek
action. Figure\,\ref{fig:diagrams} lists all diagrams which are needed
for the perturbative calculations presented in this article.

\subsection{Self-energy}
\label{app:self}

Using the expression for the vertex $V_2(p,p)$ of
eq.\,(\ref{eq:v2kw}), the tadpole contribution to the self energy can
be easily computed. In a general covariant gauge, where $\partial_\mu
A_\mu=0$, its expression is
\begin{eqnarray}
& & \frac{1}{a^2} \cdot \frac{Z_0}{2} \Big(1-\frac{1}{4}(1-\alpha)\Big) 
\cdot i a g_0^2 \CF \Big( \sum_{\mu=1}^4 \gamma_\mu a p_\mu
- \sum_{k=1}^3 \gamma_4 (1+\rmO(a^2) \Big)  \nonumber \\
&=& g_0^2 \CF \, \frac{Z_0}{2} \Big(1-\frac{1}{4}(1-\alpha)\Big) 
\, \Big( i\slash{p} - \frac{3i\gamma_4}{a} \Big) +\rmO(a),
\end{eqnarray}
where $Z_0$ is given by
\cite{GonzalezArroyo:1981ce,Ellis:1983af,Capitani:2002mp} 
\begin{equation}
  Z_0=\int_{-\pi/a}^{\pi/a} \frac{d^4p}{(2\pi)^4}
     \frac{1}{\widehat{p}^2} = 0.1549333\ldots 
     = 24.466100\,\frac{1}{16\pi^2},\qquad
     \widehat{p}^2=\frac{4}{a^2}\,\sum_\mu \sin^2\Big(\frac{ap_\mu}{2}\Big),
\end{equation}
and terms of $\rmO(a)$ and higher can eventually be set to zero. 
The term proportional to $i\slash{p}$ is the same as for Wilson
fermions, while the other term would imply a power-divergent mixing of
order~$1/a$ with the dimension-3 operator $\overline{\psi} \gamma_4
\psi$, unless there be a cancellation with an analogous term
coming from the contribution of the sunset diagrams to the self-energy
(contained e.g. in diagram (e) of Figure\,\ref{fig:diagrams}). 
Here we show that no such compensation occurs.

We have computed the sunset diagram using special computer codes
written in FORM \cite{Vermaseren:2000nd,Vermaseren:2008kw} and
Mathematica, and also checked it against calculations by hand. The
result is
\begin{eqnarray}
& &\Sigma^{sunset} (p,m_0) =
i\slash{p}\cdot\frac{g_0^2}{16\pi^2} \,\CF \,\Bigg[ \log a^2p^2 -2.99216
+(1-\alpha) \Big(-\log a^2p^2 +7.850272 \Big) \Bigg] \nonumber \\
&&\quad +\,m_0\cdot\frac{g_0^2}{16\pi^2} \,\CF \,\Bigg[ 4\,\log a^2p^2
 -24.36875 +(1-\alpha) \Big(-\log a^2p^2 +5.792010 \Big) \Bigg]  \\
&&\quad -0.12554 \cdot\frac{g_0^2}{16\pi^2} \,\CF \cdot i\, \gamma_4 p_4
 +\,(7.16687 -9.17479\,(1-\alpha))
\cdot\frac{g_0^2}{16\pi^2} \,\CF \cdot i\, \frac{\gamma_4}{a} .\nonumber
\end{eqnarray}
Note that gauge invariance forces the terms proportional to
$(1-\alpha)$ to be the same as, for example, in the case of Wilson or
overlap fermions. This is an important check of the correctness of our
calculations. The two terms proportional to $\gamma_4/a$ arising from
the tadpole and the sunset diagrams do not cancel -- they actually
reinforce each other. However, the parts proportional to $(1-\alpha)$
cancel exactly, as required by gauge invariance.

\subsection{Bilinears}
\label{app:bilin}

Here we list the results for the individual vertex diagrams for the
scalar density, as well as the vector and tensor currents. As a
consequence of chiral symmetry, the vertex correction for the
pseudoscalar density is identical to that of the scalar density. The
same is true for the vector and axial-vector currents.

For the scalar and pseudoscalar densities the result for the vertex
correction is
\begin{equation}
\Lambda_{\rm{S}} = \frac{g_0^2}{16\pi^2} \,\CF \,\Bigg[
 -4\,\log a^2p^2 + 24.36875 
 +(1-\alpha) \Big(\log a^2p^2 -5.792010 \Big) \Bigg] .
\label{eq:lambda-s}
\end{equation}
One can see that the breaking of hyper-cubic symmetry does not induce
any mixing.

For the vector current the vertex diagram yields
\begin{equation}
\Lambda_{\rm{V}}=\frac{g_0^2}{16\pi^2} \,\CF \,\gamma_\mu \Bigg[
 -\log a^2p^2 +10.44610 -2.88914\cdot\delta_{\mu 4}
+(1-\alpha) \Big(\log a^2p^2 -4.792010 \Big) \Bigg] .
\end{equation}
This result shows explicitly that the spatial and temporal components
of the vector (and also those of the axial-vector) current receive
different radiative corrections. This is a consequence of the breaking
of hyper-cubic symmetry, and of the special r\^ole taken by the
temporal direction. On the other hand, it is encouraging that mixing
between the spatial and temporal components appears to be absent. Each
of these components still renormalizes multiplicatively, and the
mixing matrix is diagonal.

Finally, for the tensor current we obtain the result for the vertex
diagram as
\begin{equation}
\Lambda_{\rm{T}}=\frac{g_0^2}{16\pi^2} \,\CF \,\sigma_{\mu\nu}\,\Bigg[
  4.17551 +(1-\alpha) \Big(\log a^2p^2 -3.792010 \Big) \Bigg] .
\end{equation}
Similarly to the scalar and pseudoscalar case, the breaking of
hyper-cubic invariance does not generate here any extra mixing. It is
remarkable that the tensor operator does not appear to show any
preference for the temporal direction even after (one-loop)
renormalization, that is, the renormalization constant is the same for
each of the six independent components of the tensor operator.

\end{appendix}


\begin{thebibliography}{10}

\bibitem{mind:Karsten81}
L.H. Karsten,
\newblock Phys. Lett. B104 (1981) 315.
\newblock 

\bibitem{mind:Wilczek87}
F. Wilczek,
\newblock Phys. Rev. Lett. 59 (1987) 2397.
\newblock 

\bibitem{mind:Creutz07}
M. Creutz,
\newblock JHEP 04 (2008) 017, arXiv:0712.1201.
\newblock 

\bibitem{mind:Borici07}
A. Bori\c{c}i,
\newblock Phys. Rev. D78 (2008) 074504, arXiv:0712.4401.
\newblock 

\bibitem{Cichy:2008gk}
K. Cichy, J. Gonzalez~Lopez, K. Jansen, A. Kujawa and A. Shindler,
\newblock Nucl. Phys. B800 (2008) 94, arXiv:0802.3637.
\newblock 

\bibitem{Bedaque:2008xs}
P.F. Bedaque, M.I. Buchoff, B.C. Tiburzi and A. Walker-Loud,
\newblock Phys. Lett. B662 (2008) 449, arXiv:0801.3361.
\newblock 

\bibitem{Bedaque:2008jm}
P.F. Bedaque, M.I. Buchoff, B.C. Tiburzi and A. Walker-Loud,
\newblock Phys. Rev. D78 (2008) 017502, arXiv:0804.1145.
\newblock 

\bibitem{Buchoff:2008ei}
M.I. Buchoff,
\newblock PoS LATTICE2008 (2008) 068, arXiv:0809.3943.
\newblock 

\bibitem{mind:Creutz08}
M. Creutz,
\newblock PoS LATTICE2008 (2008) 080, arXiv:0808.0014.
\newblock 

\bibitem{mind:Borici08}
A. Bori\c{c}i,
\newblock PoS LATTICE2008 (2008) 231, arXiv:0812.0092.
\newblock 

\bibitem{mind:Capitani09}
S. Capitani, J. Weber and H. Wittig,
\newblock Phys. Lett. B681 (2009) 105, arXiv:0907.2825.
\newblock 

\bibitem{mind:Capitani_lat09}
S. Capitani, J. Weber and H. Wittig,
\newblock (2009), arXiv:0910.2597.
\newblock 

\bibitem{Kimura:2009qe}
T. Kimura and T. Misumi,
\newblock (2009), arXiv:0907.1371.
\newblock 

\bibitem{Kimura:2009di}
T. Kimura and T. Misumi,
\newblock Prog. Theor. Phys. 123 (2010) 63, arXiv:0907.3774.
\newblock 

\bibitem{Nini1}
H.B. Nielsen and M. Ninomiya,
\newblock Nucl. Phys. B185 (1981) 20.
\newblock 

\bibitem{Nini2}
H.B. Nielsen and M. Ninomiya,
\newblock Nucl. Phys. B193 (1981) 173.
\newblock 

\bibitem{Nielsen:1981hk}
H.B. Nielsen and M. Ninomiya,
\newblock Phys. Lett. B105 (1981) 219.
\newblock 

\bibitem{impr:SW}
B. Sheikholeslami and R. Wohlert,
\newblock Nucl. Phys. B259 (1985) 572.

\bibitem{Boch}
M. Bochicchio, L. Maiani, G. Martinelli, G.C. Rossi and M. Testa,
\newblock Nucl. Phys. B262 (1985) 331.
\newblock 

\bibitem{GonzalezArroyo:1981ce}
A. Gonz\'{a}lez-Arroyo and C.P. Korthals-Altes,
\newblock Nucl. Phys. B205 (1982) 46.
\newblock 

\bibitem{Ellis:1983af}
R.K. Ellis and G. Martinelli,
\newblock Nucl. Phys. B235 (1984) 93.
\newblock 

\bibitem{Capitani:2002mp}
S. Capitani,
\newblock Phys. Rept. 382 (2003) 113, hep-lat/0211036.
\newblock 

\bibitem{Vermaseren:2000nd}
J.A.M. Vermaseren,
\newblock (2000), math-ph/0010025.
\newblock 

\bibitem{Vermaseren:2008kw}
J.A.M. Vermaseren,
\newblock Nucl. Phys. Proc. Suppl. 183 (2008) 19, arXiv:0806.4080.
\newblock 

\end{thebibliography}

\end{document}